\documentstyle[aps,prb,amsmath]{revtex}
\title{Theory of thermoelectric effects when the temperature 
approximation is incorrect}
\author{Yu.~G.~Gurevich, G.~N.~Logvinov, and O.~Yu.~Titov}
\address{Ya.~O.~Galan Ternopol' State Pedagogical Institute,\\
282009 Ternopol',  Ukraine}
\date{Submitted June 7, 1993; accepted for publication June 24, 1993}
\begin{document}
\preprint{Fiz.  Tekh.  Poluprovodn. 28, 113-119 (January 1994)}
\maketitle

\makeatletter
\global\@specialpagefalse
\def\@oddhead{%
Semiconductors {\bfseries 28} (1), January 1994, p.~68--71
\hfill
\copyright{} 1994 American Institute of Physics}
\let\@evenhead\@oddhead

\begin{abstract}
The flow of a thermoelectric current through a semiconductor of
submicron dimensions is  analyzed.  The rate of surface relaxation of
the energy is assumed to be much higher  than the rate of
electron-electron collisions.  Under these conditions, it is incorrect
to describe  the electron gas by means of a Maxwellian distribution and
thus to describe the  thermoelectric effects in terms of an electron
temperature and a chemical potential.  A theory  is derived for these
effects.  This theory does not include the latter parameters and is 
based on a non-Maxwellian distribution which is spatially nonuniform in
terms of energy.
\end{abstract}
\pacs{pacs here}

In the conventional theories on thermoelectric effects,
the temperature and the chemical potential are fundamental 
parameters used by the theoretician.  As was shown in
Ref.~\onlinecite{GLT93}, on the other hand, when a submicron-size 
semiconductor layer
makes contact with a heater and a refrigerator, the symmetric part of
the distribution function of the
current carriers is essentially non-Maxwellian (non-Fermi), so such
concepts as temperatures and chemical
potentials are not at all valid.\cite{foot1} There is accordingly a need
to derive a theory of thermoelectric effects which is not
based on these concepts.  In the present paper we attempt
to derive such a theory.

For definiteness we consider a model of a closed thermoelectric circuit
consisting of a semiconductor layer of thickness $2a$, an external
metallic region of length $L$, and transition regions (contacts) with
thicknesses $2\delta\to 0$ between the semiconducting and metallic
media.\cite{BBG84,GL92} We choose the thickness $2a$ to be so small that
the inequality $2a\ll l_{\varepsilon}$ holds (we are thus dealing with 
a submicron-size layer), where $l_{\varepsilon}$, is the diffusion 
length for the relaxation of the energy of electrons among scattering
centers.  For quasielastic scattering events, this length is on the
order of 1--10 $\mu$m.  The plate surfaces $x= \mp(a + \delta)$ are in 
contact with heat reservoirs at temperatures $T_{1}$ and $T_{2}$
($T_{1} > T_{2}$).  As a result, there are spatial variations in the
distribution function, and a voltage is generated in the circuit.  The
impact of these variations increases with increasing frequency of the
surface energy relaxation, $\nu_{s}$.  Since the latter is inversely
proportional to the layer thicknesses\cite{GL90} the rate of 
electron-electron collisions in a submicron layer at a fixed equilibrium
density may be much lower than $\nu_{s}$. The electron gas would
therefore not be correctly described by means of a Maxwellian
distribution function with an electron temperature $T_{e}$.  In this 
case the symmetric part of the distribution function should be
determined from a Boltzmann kinetic equation.

As the mechanism for the inelastic interaction of the nondegenerate
electron gas with the heat reservoirs we adopt a mechanism involving the
flow of a ``partial'' current, by which we mean a current of a group of
electrons with a given energy, through surface layers.  Obviously,
energy may be dissipated both at centers within the surface layers and
in the course of a direct interaction with the heat reservoirs.  Since
our purpose here is not to derive exact quantitative results, but
instead to establish fundamental positions of this theory, the energy
dissipation within the surface layers is not of fundamental importance
and it can be ignored.\cite{GLT93}

The symmetric part of the electron distribution function calculated
under these assumptions, in the approximation linear in the parameter
$\Delta T/T\ll 1$ [$\Delta T= T_{1} - T_{2}$, $T = (T_{1} - T_{2})/2$],
is\cite{GLT93}
\begin{equation}
f_{0}(\varepsilon,x) = e^{(\mu - \varepsilon)/T}
\left[
1 + \Psi(\varepsilon)\frac{x}{a}\frac{\Delta T}{T}
\right] .
\label{eq:glt94:f0}
\end{equation}
Here
\begin{equation}
\Psi(\varepsilon, x) = A(\varepsilon)M(\varepsilon) + 
P(\varepsilon)j_{0} + C_{1} Q(\varepsilon) + C_{2} R(\varepsilon) ,
\label{eq:glt94:psi}
\end{equation}
where
\[
A(\varepsilon) = 
\frac{a\xi_{s}(\varepsilon)}{\xi(\varepsilon) + a\xi_{s}(\varepsilon)};
\quad
P(\varepsilon) =
\frac{a\Lambda(\varepsilon)}{\xi(\varepsilon) + a\xi_{s}(\varepsilon)};
\]
\[
Q(\varepsilon) =
\frac{a\sigma(\varepsilon)}{\xi(\varepsilon) + a\xi_{s}(\varepsilon)};
\quad
R(\varepsilon) =
\frac{a\sigma_{s}(\varepsilon)}{\xi(\varepsilon) + 
a\xi_{s}(\varepsilon)};
\]
\[
\Lambda(\varepsilon) =
\frac{\sigma(\varepsilon)}{\sigma} - 
\frac{\sigma_{s}(\varepsilon)}{\sigma_{s}} ;
\quad
M(\varepsilon) =
\frac{3}{4} - \frac{\varepsilon}{2T};
\]
\[
\xi(\varepsilon) =
\frac{2e\varepsilon g(\varepsilon)\tau(\varepsilon)
e^{(\mu - \varepsilon)/T}}{3m}; 
\quad
\sigma(\varepsilon) =
\frac{2e^{2}\varepsilon g(\varepsilon)\tau(\varepsilon) 
e^{(\mu - \varepsilon)/T}}{3mT};
\]
\[
\xi_{s}(\varepsilon) = \lim_{\delta \to 0}
\frac{2e\varepsilon g_{s}(\varepsilon)\tau_{s}(\varepsilon)
e^{(\mu - \varepsilon)/T}}{3m\delta}; 
\quad
\sigma_{s}(\varepsilon) =\lim_{\delta \to 0}
\frac{2e^{2}\varepsilon g_{s}(\varepsilon)\tau_{s}(\varepsilon) 
e^{(\mu - \varepsilon)/T}}{3mT\delta};
\]
$j_{0} = j/(\Delta T/T)$; $j$ is the thermoelectric current density; 
$\sigma = \int_{0}^{\infty}d\varepsilon\,\sigma(\varepsilon)$ is the 
bulk electrical conductivity; $e$, $m$, and $\varepsilon$ are the 
charge, effective mass, and energy of an electron;
$\mu$ is the chemical potential at the 
temperature\cite{foot2} $T$; $\mu(\varepsilon)$ is the 
density of electron states; $\tau(\varepsilon)$ is the momentum
relaxation time; and 
\[
C_{1} = \frac{1}{a\sigma}\int_{0}^{\infty}d\varepsilon 
\, \xi(\varepsilon) \Psi(\varepsilon)
\]
and $C_{2}$ are constants to be determined.  The subscript $s$ refers to
surface layers.

It follows from these definitions that the functions
$\sigma(\varepsilon)$ and $\sigma_{s}(\varepsilon)$ constitute partial 
bulk and surface conductivities.  We have $\xi(\varepsilon) = 
(T/e)\sigma(\varepsilon)$. For a surface layer, we cannot, in general,
write a similar relation for $\xi_{s}(\varepsilon)$ and 
$\sigma_{s}(\varepsilon)$ at the outset, since the kinetic coefficients 
of the boundary layer change with decreasing thickness $2\delta$, and
they may acquire singularities\cite{BBG84} in the limit $2\delta \to 0$. 
It is not possible to follow these changes in the absence of a 
microscopic theory for surface relaxation processes.  We will
accordingly assume for simplicity $\xi_{s}(\varepsilon) = 
(T/e)\beta_{s}\sigma_{s}(\varepsilon)$,
where $\beta_{s}$ is a phenomenological surface parameter with the
dimensionality of a reciprocal length.  We do not believe that this
simplification is of fundamental importance.  The important point is
that there is no relationship of any sort between the functions 
$\sigma(\varepsilon)$ and $\sigma_{s}(\varepsilon)$.

Under this assumption we can rewrite the functions in
$\Psi(\varepsilon)$ as follows:
\begin{equation}
\begin{gathered}
A(\varepsilon) =
\frac{Z(\varepsilon)}{1 + Z(\varepsilon)},
\quad
P(\varepsilon) =
\frac{ea}{T}\frac{\Lambda(\varepsilon)}
{\sigma(\varepsilon)\left[1 + Z(\varepsilon)\right]} ,\\
Q(\varepsilon) =
\frac{ea}{T}\frac{1}{1 + Z(\varepsilon)},
\quad
R(\varepsilon) =
\frac{e}{\beta_{s}T}\frac{Z(\varepsilon)}{1 + Z(\varepsilon)} ,
\end{gathered}
\label{eq:glt94:APQR}
\end{equation}
where
\begin{equation}
Z(\varepsilon) = 
\frac{a\beta_{s}\sigma_{s}(\varepsilon)}{\sigma(\varepsilon)} .
\end{equation}

The first term in Eq.~(\ref{eq:glt94:psi}) corresponds to the effect of 
the heat reservoirs on the distribution function in the case in
which these reservoirs are related to the electron gas of the
submicron layer by the part of the partial currents caused
by the internal thermoelectric field and by thermal diffusion [in the
temperature approximation, $\Psi(\varepsilon) = M(\varepsilon) = 3/4 -
\varepsilon/2T$; Ref.~\onlinecite{GLT93}]. The second term reflects the 
mechanism for the formation of a non-Maxwellian distribution function
involving the flow of a macroscopic current across the interface between
two media differing in the energy dependence of the relaxation 
time.\cite{GLT93} The last two terms stem from the drift components of
the partial current in the semiconductor and in the transition layer.

To find the constants $C_{1}$ and $C_2$, we use the definition
of the coefficient $C_{1}$ and the normalization condition on the
function $\Psi(\varepsilon)$ (Ref.~\onlinecite{GLT93}):
\begin{equation}
\int_{0}^{\infty}
d\varepsilon\,\sqrt{\varepsilon}\Psi(\varepsilon)e^{-\varepsilon/T} = 0 
.
\label{eq:glt94:norm}
\end{equation}
As a result, we find the system of algebraic equations
\begin{equation}
\begin{gathered}
\gamma_{1}C_{1} + \gamma_{2}C_{2} = - \gamma_{3} - \gamma_{4}j_{0} ,\\
\gamma'_{1}C_{1} - \gamma'_{2}C_{2} = \gamma'_{3} + \gamma'_{4}j_{0} .
\end{gathered}
\label{eq:glt94:gammaeqs}
\end{equation}
Here
\begin{equation}
\begin{gathered}
\gamma_{1} = 
\int_{0}^{\infty}d\varepsilon\,\eta(\varepsilon)Q(\varepsilon) ,\\
\gamma'_{1} = 
1 - (T/ea\sigma)\int_{0}^{\infty}d\varepsilon\,\sigma(\varepsilon)
Q(\varepsilon) ,\\
\gamma_{2} = 
\int_{0}^{\infty}d\varepsilon\,\eta(\varepsilon)R(\varepsilon), 
\quad
\gamma'_{2} = 
(T/ea\sigma)\int_{0}^{\infty}d\varepsilon\,\sigma(\varepsilon)
R(\varepsilon) ,\\
\gamma_{3} = 
\int_{0}^{\infty}d\varepsilon\eta(\varepsilon)A(\varepsilon)
M(\varepsilon) ,\\
\gamma'_{3} = (T/ea\sigma)
\int_{0}^{\infty}d\varepsilon\sigma(\varepsilon)A(\varepsilon)
M(\varepsilon) ,\\
\gamma_{4} = 
\int_{0}^{\infty}d\varepsilon\,\eta(\varepsilon)P(\varepsilon),
\quad
\gamma'_{4} = 
(T/ea\sigma)\int_{0}^{\infty}d\varepsilon\sigma(\varepsilon)
P(\varepsilon) 
\end{gathered}
\label{eq:glt94:gammas}
\end{equation}
are constant, where $\eta(\varepsilon) = 
\sqrt{\varepsilon}e^{-\varepsilon/T}$.

Determining the coefficients $C_{1}$ and $C_{2}$ from system
of Eqs.~(\ref{eq:glt94:gammaeqs}), and substituting the results into 
Eq.~(\ref{eq:glt94:norm}) and~(\ref{eq:glt94:f0}), we find an expression 
for the symmetric part of the electron distribution function in the
submicron layer:
\begin{equation}
f_{0}(\varepsilon, x) = e^{(\mu - \varepsilon)/T}
\left\{
1 + \frac{x}{a}
\left[
F(\varepsilon) + G(\varepsilon) j_{0}
\right]
\frac{\Delta T}{T}
\right\} ,
\label{eq:glt94:f0layer}
\end{equation}
where
\begin{equation}
F(\varepsilon) = A(\varepsilon)M(\varepsilon) + \Gamma_{1}Q(\varepsilon)
+ \Gamma_{3}R(\varepsilon) ,
\label{eq:glt94:F}
\end{equation}
\begin{equation}
G(\varepsilon) = P(\varepsilon) + \Gamma_{2} Q(\varepsilon) + \Gamma_{4}
R(\varepsilon), 
\label{eq:glt94:G}
\end{equation}
\begin{equation}
\begin{gathered}
\Gamma_{1} = \frac{\gamma_{2}\gamma'_{3} - \gamma'_{2}\gamma_{3}}
{\gamma_{1}\gamma'_{2} - \gamma'_{1}\gamma_{2}},
\quad
\Gamma_{2} = \frac{\gamma_{2}\gamma'_{4} - \gamma'_{2}\gamma_{4}}
{\gamma_{1}\gamma'_{2} - \gamma'_{1}\gamma_{2}}, \\
\Gamma_{3} = \frac{\gamma_{1}\gamma'_{3} - \gamma'_{1}\gamma_{3}}
{\gamma_{1}\gamma'_{2} - \gamma'_{1}\gamma_{2}},
\quad
\Gamma_{4} = \frac{\gamma_{1}\gamma'_{4} - \gamma'_{1}\gamma_{4}}
{\gamma_{1}\gamma'_{2} - \gamma'_{1}\gamma_{2}} .
\end{gathered}
\label{eq:glt94:Gammas}
\end{equation}

The thermal emf generated in the closed circuit is
found from the expression\cite{GY91}
\begin{equation}
E = jR = \oint\frac{j}{\sigma}\, dx = \lim_{\delta to 0}
\left(
\int_{-a + \delta}^{ a - \delta}\frac{j}{\sigma}\, dx +
\int_{ a - \delta}^{ a + \delta}\frac{j}{\sigma}\, dx +
\int_{ a + \delta}^{-a - \delta}\frac{j}{\sigma}\, dx +
\int_{-a - \delta}^{-a + \delta}\frac{j}{\sigma}\, dx
\right) .
\label{eq:glt94:E}
\end{equation}
Here $R = R_{s} + R_{c}+ R_{L}$ is the total ohmic resistance of the
closed circuit, $R_{s} = 2a/\sigma$, is the resistance of the surface
layer, $R_{c} = \lim_{\delta to 0}
[2\delta/\int_{0}^{\infty}d\varepsilon\,\sigma_{s}(\varepsilon)]$
is the resistance of the contacts, and $R_{L}=L/\sigma_{L}$ is the 
resistance of the external part
of the circuit, whose conductivity $\sigma_{L}$.

For simplicity we set the cross-sectional area of the
contour equal to one everywhere.

The first and third integrals in Eq.~(\ref{eq:glt94:E}) correspond to the
components of the thermal emf due to the semiconductor
and the external region.  The second and fourth integrals
determine the thermal emf in the contacts.

The electric current density $j$ can be expressed in
terms of the symmetric part of the distribution function:
\begin{equation}
j = \int_{0}^{\infty}d\varepsilon\, j(\varepsilon, x) ,
\label{eq:glt94:j}
\end{equation}
where the partial current $j(\varepsilon, x)$ is\cite{RGK76}
\begin{equation}
j(\varepsilon, x) = - 
\frac{2e\varepsilon g(\varepsilon)\tau(\varepsilon)}{3m} 
\left(
\frac{\partial f_{0}}{\partial x} + eE\frac{\partial 
f_{0}}{\partial\varepsilon} 
\right) ,
\end{equation}
$E = -d\varphi/dx$ is the electric field, and $\varphi$ is the electric
potential.

If we ignore the contribution to the thermal emf in the
metallic part of the circuit, the integration over the closed
contour in Eq.~(\ref{eq:glt94:E}) reduces to
\begin{multline}
jR = - \frac{2\Delta T}{e\sigma} 
\int_{0}^{\infty}d\varepsilon\,\sigma(\varepsilon)F(\varepsilon) -
j\frac{2T}{e\sigma}\int_{0}^{\infty} 
d\varepsilon\,\sigma(\varepsilon)G(\varepsilon) \\
- \frac{T}{e\sigma_{s}} 
\int_{0}^{\infty}d\varepsilon\,\sigma_{s}(\varepsilon) 
e^{(\varepsilon - \mu)/T} 
\left[
f_{00}(\varepsilon, x = a) - f_{0}(\varepsilon, x = a) -
f_{00}(\varepsilon, x = -a) + f_{0}(\varepsilon, x = -a)
\right] .
\label{eq:glt94:jR}
\end{multline}
Here $f_{00}(\varepsilon, x = \pm a)$ are Fermi distribution functions 
for the electrons in the metal, with temperatures $T_{2}$ and $T_{1}$,
respectively.

Setting 
$f_{00}(\varepsilon, x = a) \approx f_{00}(\varepsilon, x = -a)$,
and carrying out some simple calculations, we find an expression for the
thermoelectric current $j$:
\begin{equation}
j =
\frac
{\frac{2}{e} \int_{0}^{\infty}d\varepsilon
\left[
\frac{\sigma_{s}(\varepsilon)}{\sigma_{s}} -
\frac{\sigma(\varepsilon)}{\sigma}
\right]
F(\varepsilon)}
{R + \frac{2T}{e}\int_{0}^{\infty}d\varepsilon
\left[
\frac{\sigma(\varepsilon)}{\sigma} -
\frac{\sigma_{s}(\varepsilon)}{\sigma_{s}}
\right]
G(\varepsilon)}
\Delta T .
\label{eq:glt94:jDT}
\end{equation}

Since the function $\Lambda(\varepsilon) = \sigma(\varepsilon)/\sigma - 
\sigma_{s}(\varepsilon)/\sigma_{s}$
appears linearly in $G(\varepsilon)$ [see Eq.~(\ref{eq:glt94:G}) 
and~(\ref{eq:glt94:Gammas})], the denominator of 
Eq.~(\ref{eq:glt94:jDT}) has no singularities.

After we substitute Eq.~(\ref{eq:glt94:jDT}) into 
Eq.~(\ref{eq:glt94:E}).
we can write the thermal emf in the customary form
\begin{equation}
E = a\Delta T
\label{eq:glt94:EDT}
\end{equation}
where the thermal-emf coefficient $a$ is
\begin{equation}
a = 
\frac
{\frac{2}{e} \int_{0}^{\infty}d\varepsilon
\left[
\frac{\sigma_{s}(\varepsilon)}{\sigma_{s}} -
\frac{\sigma(\varepsilon)}{\sigma}
\right]
F(\varepsilon)}
{R + \frac{2T}{e}\int_{0}^{\infty}d\varepsilon
\left[
\frac{\sigma(\varepsilon)}{\sigma} -
\frac{\sigma_{s}(\varepsilon)}{\sigma_{s}}
\right]
G(\varepsilon)}
R
\label{eq:glt94:a}
\end{equation}

If the electron gas in the semiconductor is Maxwellized, the expression
for the thermal-emf coefficient becomes\cite{GL92}
\begin{equation}
a = 
\frac{(1 - \beta)a_{s} + \beta a_{v}}{R + R_{P}} R ,
\label{eq:glt94:aMax}
\end{equation}
where the coefficient $\beta$ determines the extent to which the
thermal contact is isothermal, $a_{v}$ and $a_{s}$ are the bulk and
surface components of the thermal emf, respectively, and
$R_{P}$ is the Peltier resistance.

It can be concluded from a comparison of Eq.~(\ref{eq:glt94:a}) 
and~(\ref{eq:glt94:aMax}) that the quantity
\[
R' = \frac{2T}{e} \int_{0}^{\infty}d\varepsilon
\left[
\frac{\sigma(\varepsilon)}{\sigma} -
\frac{\sigma_{s}(\varepsilon)}{\sigma_{s}} 
\right]
G(\varepsilon)
\]
is the Peltier resistance for the conduction electrons described by
distribution Eq.~(\ref{eq:glt94:f0}).  If $\sigma(\varepsilon) 
=\sigma_{s}(\varepsilon)$, then the thermal current and thermal emf
vanish according to Eqs.~(\ref{eq:glt94:jDT}) and~(\ref{eq:glt94:a}). 
The evident reason for this vanishing is that the bulk and surface
components of the overall thermal emf cancel out under the assumption
that the partial conductivities $\sigma(\varepsilon)$ and 
$\sigma_{s}(\varepsilon)$ are proportional to, respectively, the
parameters $\xi(\varepsilon)$ and $\xi_{s}(\varepsilon)$, which are 
responsible for the thermal coupling of the electrons of the submicron
layer with the heat reservoirs.\cite{GLT93} In this case, the
Peltier resistance clearly must vanish automatically.

Let us consider two particular cases corresponding to
extreme values of the function ${\cal Z}(\varepsilon)$: 
${\cal Z}(\varepsilon) \ll 1$ and ${\cal Z}(\varepsilon) \gg 1$.  It is 
easy to show, on the basis of Eqs.~(\ref{eq:glt94:APQR}) 
and~(\ref{eq:glt94:gammas}), that we have $F(\varepsilon) to 0$ in the 
first case; this circumstance means that there is no thermal current
in the circuit. In this case, distribution function 
Eq.~(\ref{eq:glt94:f0}) transforms into a Maxwellian equilibrium
distribution function with an equilibrium temperature $T$. This 
situation could naturally be called ``adiabatic.'' Rewriting the
inequality ${\cal Z}(\varepsilon) \ll 1$ as
$\sigma(\varepsilon)/a \gg \beta_{s}\sigma_{s}(\varepsilon)$, we see 
that, for given values of the surface parameters, the ``adiabaticity''
of the system becomes more pronounced with thinner samples, i.e., with
increasing partial conductivity of the semiconductor layer per unit
length.  This situation is equivalent to an effective decrease
in the surface conductivity and thus a weakening of the
coupling of the carriers with the heat reservoirs.  In the
temperature approximation, this situation corresponds to
the condition that the thermal conductivity per unit length
is greater than the surface thermal conductivity.\cite{BBG84}          
In this case, the temperature $T_{e}$, becomes uniform along the
length of the sample because of the large bulk thermal
conductivity.

In the second limiting case [${\cal Z}(\varepsilon) \gg 1$] we have 
$F(\varepsilon) = M(\varepsilon)$, $G(\varepsilon) \to 0$, and the 
thermal-emf coefficient is
\begin{equation}
a = \frac{1}{e}
\left[
-
\frac{\frac{1}{\sigma_{s}} 
\int_{0}^{\infty}d\varepsilon\,\sigma_{s}(\varepsilon)}{T} +
r + \frac{5}{2}
\right] ,
\label{eq:glt94:ar}
\end{equation}
where the number $r$ determines the relaxation time of the
momentum of the electrons in the semiconductor:
$\tau(\varepsilon) = \tau_{0}(\varepsilon/T)^{r}$.

In this case the distribution function (\ref{eq:glt94:f0}) becomes a
Maxwellian distribution corresponding to the temperature
\begin{equation}
T_{e}(x) = T - \frac{x}{2a} \frac{\Delta T}{T} .
\label{eq:glt94:Te}
\end{equation}

This result may look paradoxical, as contradicting our
original assumption that it is incorrect to describe the electrons by
means of a Maxwellian distribution function.  In
speaking of a ``Maxwellization'' of a gas of carriers one is
usually assuming that a dominant role is played by an
electron-electron collision integral; such an integral can be
omitted from the problem at hand.  Consequently, the conventional
mechanism for the shaping of a Maxwellian distribution function is
actually meaningless.  In submicron layers, however, one can point out
another mechanism for a Maxwellization of a gas: the intense ``mixing'' 
of the partial electron current densities as a result of their
interaction with the heat reservoirs.  The heat reservoirs, in fact,
determine the temperature (\ref{eq:glt94:Te}) of the electron gas.
Clearly, as the frequency $\nu_{v}$, decreases, this mixing weakens,
and the distribution function becomes non-Maxwellian.  It
is also clear that the boundary conditions should be quite
different from adiabatic conditions in this case, since otherwise the
thermoelectric effects would vanish.  The boundary conditions describing
the second limiting case can be
found as isothermal boundary conditions which lead to an
ideal coupling of the electrons with the heat reservoirs.

In the temperature approximation, the thermal emf is
usually found from\cite{A78}
\begin{equation}
E = \frac{1}{e} 
\left(
 - \frac{\mu}{T} + r + \frac{5}{2}
\right)
\Delta T
\label{eq:glt94:Eta}
\end{equation}
It contains a thermal-diffusion component $E_{T} = (1/e) (r + 1) \Delta 
T$ and a contribution from the contact
potential difference, $E_{C} = (1/e)(3/2 - \mu/T)\Delta T$.  Under
nonisothermal boundary conditions, a third component is
added: the surface thermal emf in (\ref{eq:glt94:aMax}).  This 
partitioning of the thermal emf into components can be carried out
because physical quantities---the chemical potential and
the temperature---are distinguished in the Maxwellian distribution
function.  The temperature plays a twofold role in
thermoelectric phenomena.  First, the temperature gradient
determines the ``external force'' acting on the carriers.  Second, the
same gradient specifies the spatial variation of the
chemical potential\cite{foot3} and thus the internal thermoelectric
field.

Comparing Eq.~(\ref{eq:glt94:ar}) and~(\ref{eq:glt94:Eta}), we see that 
in submicron semiconductors under isothermal boundary conditions it is
again possible to separate contributions from thermal diffusion atid
contact components of the thermal emf.  Furthermore, the contact emf can
be determined from an expression which includes exclusively surface
characteristics:
\begin{equation}
E_{C} = \frac{1}{e}
\left[
\frac{3}{2} - 
\frac{\frac{1}{\sigma_{s}}
\int_{0}^{\infty}d\varepsilon\,\varepsilon\sigma_{s}(\varepsilon)}
{T} 
\right]
\Delta T .
\end{equation}

Under nonisothermal boundary conditions, it is no
longer possible to partition the thermal emf into bulk, contact, and
surface components, as can be seen from expression~(\ref{eq:glt94:a}). 
All we can do is speak in terms of ``bulk'' and
``surface'' components of the thermal emf, without subdividing the
latter into ``contact'' and ``thermal-diffusion''
parts.

The two limiting cases discussed above can occur only
if the inequalities written above for the function 
${\cal Z}(\varepsilon)$
hold for all or at least most of the electrons.  If the energy
dependence of the relaxation time in the interior is different
from that in the surface layer, then (for example) adiabatic boundary
conditions may hold for the electrons of
one group, and isothermal boundary conditions for those
of another.  The behavior of the overall electron gas will
then be determined by an effective resultant contribution of
each of these groups.

Translated by D. Parsons


\begin{references}
\bibitem{foot1}
There are several other physical situations (Ref.~\onlinecite{BS84}, for
example) which require a study of thermoelectric effects in the absence
of a temperature and a chemical potential.
\bibitem{foot2}
For simplicity we assume that the chemical potential has the same value 
everywhere in the equilibrium state.
\bibitem{foot3}
In inhomogeneous semiconductors, the $\mu(x)$ dependence may be governed
by other factors also, e.g., a nonuniform doping.
\end{references}
\end{document}